\documentstyle [aps]{revtex}
\begin{document}                                                    
\draft                                                      
\title{ Colossal Magnetoresistance using the
 Small Polaron Picture with Finite Bandwidth Effects}
\author{Sudhakar Yarlagadda}
\address{Saha Institute of Nuclear Physics, Calcutta, India \\
and Texas Center for Superconductivity, 
Univ. of Houston, TX 77204-5932}
\date{\today}
\maketitle

\begin{abstract}
 We present a small polaron picture and show that finite bandwidth
effects are important to understand colossal magnetoresistance.
Besides the polaron size parameter, we show that there is another
parameter (adiabaticity parameter) that is relevant
to studying magnetoresistance. We find that
for fixed values of the polaron size parameter an increase in the
adiabaticity parameter increases the magnetoresistance.
The magnetic transition is  studied within a mean field approach.
We point out important oversights in the literature.
We find that for the reported values of the bandwidth
(based on band structure calculations) and for experimentally
determined values of activation energy and Debye frequency,
the calculated values of the magnetoresistance 
compare favorably with experimental ones.
We calculate the optical conductivity too and find that
there is reasonable
 agreement with experiment.
\end{abstract}


                                                                  
                                                                   
\section{INTRODUCTION}                                
Studying perovskite manganites
of the form $\rm A_{1-\delta}B_{\delta} MnO_3$
 (A=La, Pr, Nd, etc.; B=Sr, Ca, Ba, etc.)
as a function of doping $\delta$
has lead to a variety of rich phenomena.\cite{ramirez,bishop,khomsawa}
Of these  $\rm La_{1-\delta}Ca_{\delta} MnO_3$
is perhaps the simplest one because the ionic size
difference between $ \rm La^{3+}$ and $ \rm Ca^{2+}$ is less than
$3\%$.
In $ \rm La_{1-\delta}Ca_{\delta}MnO_3$, 
at low doping, as temperature is lowered the system 
undergoes orbital ordering and 
at even lower temperatures a layered
antiferromagnetism is observed.\cite{allen} 
 At intermediate doping
($\delta \sim 0.2-0.4$),
 simultaneous metal-insulator (MI)
 and paramagnetic-ferromagnetic
transitions occur in this compound as the temperature is decreased.
At even higher doping
 (i.e., greater than $\delta \sim$ 0.5)
 charge ordering is realized 
while at $\delta \sim 1$ antiferromagnetic order results
at low temperatures.
\cite{ramirez}
To explain the magnetic ordering at low doping,
 de Gennes \cite{degennes} some time
ago had proposed double exchange
mechanism wherein, on account of strong Hund's coupling
between the spin of a mobile hole and the spin of the localized
electrons, the hopping integral of the itinerant hole
is reduced by half of the cosine of the angle between the
$3/2$ spins of the localized electrons on neighboring sites.
 Furukawa pioneered in demonstrating the usefulness
of the dynamical mean field theory in understanding the
properties of double exchange systems.\cite{furukawa}
 However it was recognized by Millis et al. \cite{millis1} 
that double exchange mechanism itself is not sufficient
to explain colossal magnetoresistance (CMR). Millis and co-workers have
proposed a model \cite{millis2} which uses Jahn-Teller coupling
 between electrons and nuclei. 
However this model treats phonons classically and 
 does not seem to yield satisfactory results
away from half-filling.
 Furthermore, these authors have studied the
phenomenon using only the polaron size parameter
 (the ratio between the hopping integral
and the binding energy). They have not studied the
effect of another
dimensionless parameter -- the ratio between the hopping integral and
the Debye frequency (adiabaticity parameter).
 R\"{o}der et al. have also stressed the
importance of Jahn-Teller coupling
in understanding these manganites.\cite{roder} Lee and Min
too have studied polaron transport in manganites.\cite{lee}
However these authors do not take into account the renormalization
of the electron-phonon interaction due to finite bandwidth effects.
 Jaime et al. \cite{jaime} and
 Worledge et al. \cite{worledge}
 have demonstrated that
their high temperature resistivity data fits well to an adiabatic small
polaron model. All in all there is growing evidence for a 
small polaron picture to explain CMR.

In this paper we study CMR 
phenomena in perovskite manganites 
by considering the 
carriers as small polarons whose high temperature 
behavior is hopping type and low temperature behavior 
is metal-like.\cite{ting}
 Our model includes effects due to electron-phonon
coupling and on-site Hund's
coupling between itinerant holes and localized electrons.
To understand
MI transition we simplify the hamiltonian by
accounting for the Hund's coupling through the double exchange
hopping term.\cite{degennes}
By including finite bandwidth effects, and using a suggestion
by Toyozawa \cite{toyozawa}
we obtain an expression for the small polaronic
wavefunction. Using this {\it nearly} small polaronic wavefunction
we obtain the dynamic conductivity.
We find that in the presence of a magnetic field
 both double exchange and finite bandwidth effects 
lower the resistivity and shift its peak to higher
temperatures and thus can lead to CMR.
In our picture the main reason for the CMR is due to the
renormalization of the electron-phonon interaction (or the
lattice distortion) due
to finite bandwidth effects.
One of our important conclusions is that for a fixed
value of the polaron size parameter the magnetoresistance increases as 
the adiabaticity parameter increases.

Within a mean-field approach, we calculate the magnetization ($M$)
of the localized spins.
The magnetization $M$ is a result of
 the effective magnetic field generated by the band like
motion of the electrons or in other words, the itinerant electrons
 due to the
 strong Hund's coupling polarize the localized spins.
We have studied the magnetization
for both with and without external magnetic fields
and found that our M values are qualitatively in
agreement with experimental results.\cite{urushibara}
 Furthermore our magnetoresistance values
also compare favorably with experimental ones.\cite{urushibara}

We also calculate the optical conductivity above the metal-insulator
transition temperature. We find that the optical conductivity
scaled by the DC conductivity depends only on the renormalized 
electron-phonon coupling and the Debye frequency. We calculate
optical conductivity as a function of frequency at various temperatures
and find reasonable agreement with experiments.\cite{quijada}

\section{BACKGROUND}

Earlier on Zener proposed a double exchange model where conduction
from a $Mn ^{3+}$ to a $Mn^{4+}$ via an $O^{2-}$ can be understood
in terms of simultaneous transfer of two electrons --  one electron from
a $Mn ^{3+}$ to a neighboring $O^{2-}$ and another electron 
from  the $O^{2-}$ to a neighboring $Mn^{4+}$. De Gennes further
realized that since the Hund's coupling is much larger than the
hopping integral $t$, the hopping term gets modified
to be $t \cos (\theta /2)$ where $\theta$ is the angle
between neighboring spins that are treated classically.
More recently it was pointed out that
there is a Berry phase factor that enters the hopping term.
\cite{hartman}
However, we feel that the geometric phase
will not significantly alter the basic Physics behind CMR.
Then to understand propagation of holes, of doping level $\delta$,
along a chain of spins (all with spin $S$)
coupled antiferromagnetically through the coupling parameter $J$
one needs to minimize the energy $E = -2t\delta \cos (\theta /2)
+ |J| S^2 \cos (\theta )$  to obtain the ground state canting
angle $\cos (\theta /2 ) = t\delta /(2 |J| S^2)$.

Within a mean-field approach one can approximate $\cos (\theta )$
to be given by $\langle \vec{S}_i /S\rangle
\cdot \langle \vec{S}_j/S \rangle =
 M^2 /M_S ^2$ where $M/M_S$ is the scaled magnetization.
 Thus the hopping term gets modified
to be $t \sqrt{(1+M^2/M_S^2)/2}$. However this double 
exchange modification of the hopping integral
 itself does not explain the
the observed metal-insulator
transition in the intermediate doping regime. 

To understand 
the colossal magnetoresistance phenomenon 
we will now use small polaronic picture.
 First we will provide
the motivation for 
this approach.
One of the striking features of  a single small polaron is that
the shape of the inverse of the mobility as a function of temperature
is quite similar to that of the resistivity observed for the manganite
 systems that display colossal magnetoresistance.
To understand how this comes about
we will present our understanding of a small polaron first
by ignoring spin effects.
 In systems like the transition metal
oxides the electron couples to the vibrational modes
of the host molecule (say the breathing mode).
Due to strong electron-phonon coupling the molecular
equilibrium configuration gets distorted. The electron 
 gets bound  in the distortion to form a polaron. This
 composite entity, i.e.,  the electron plus distortion, is the polaron.
When the distorted region is less than a lattice spacing
(i.e., 1/2 bandwidth $<$ binding energy) we have a small polaron.
The polaron propagates just like an electron.
However, the hopping integral gets modified because one has to
take into account the
wave functions of the host molecules 
which correspond to 
displaced simple harmonic oscillators.
 The overlap between the simple harmonic oscillator
 wavefunctions between adjacent sites decreases with increasing
temperature because as temperature increases higher eigenfunctions
with more nodes come into play.
The band energy thus assumes the form
 $\epsilon _{k} \sim -2t ( \cos (k_x a) + \cos (k_y a) + ..)
\exp[- g_{0}^2 \coth (\omega _0 /2 k_B T)] $ with $g_0$
being proportional to molecular distortion and $\omega _0$
being the Debye energy. As can be seen the band energy
 at high temperatures decreases exponentially with temperature.

 We are dealing with narrow band systems that are non-degenerate and
hence the mobility is given by $\mu =eD/(k_B T)$ where $D$ is the
diffusivity. At higher temperatures band narrowing occurs
and when coherent motion is no longer possible
the electron becomes localized and propagates by hopping.
The diffusivity is then given by $D \sim \frac{a^2}{\tau}$
where the life time $\tau$,  for adiabatic transport,
is given by the classical result
$1/\tau = \omega _0 /(2 \pi ) \exp [-E_a /(k_B T)]$ with
the activation energy $E_a = 2 g_0^2 \tanh [\omega _0 /(4 k_B T)]$
 (see Holstein's article \cite{holstein} for details). 
On the other hand at low temperatures band-like
conduction is possible and the diffusivity is given by
$D \propto \langle v ^2 \rangle \tau$ with $v$ being the velocity
and  $\tau$ being  still given by the above formula.
The crossover from band-like motion to hopping conductivity
takes place when the uncertainty in energy ($\hbar / \tau  $)
is of the order of half the bandwidth.
Now, it is important to point out that the total mobility 
is the sum of the band and hopping mobilities 
(see Friedman's paper \cite{friedman} for details).
The total mobility is then given by 
 \begin{eqnarray} 
\mu _{Total} =
 \beta q_e a^2 && \left [  2 \pi  \frac{t^2}{\omega_{0} }
 \exp {[-2 g_0^2 
 {\rm csch} (\frac{\beta \omega_0}{2})]}  \right . 
 \nonumber \\                           
  && 
\left . 
+  \frac{\omega _0}{2 \pi}
  \exp{[-2 g_0^2 \tanh (\frac{\beta \omega_0}{4})]} \right ],
\label{rho}
 \end{eqnarray} 
where $q_e$ is the electronic charge and $\beta = 1/(k_{B}T)$.

When we include Hund's coupling,
$t$ gets modified to be $t \sqrt{(1+M^2/M_S^2)/2}$. 
Now when a magnetic field gets switched on, the 
value of the effective
hopping integral increases, lattice gets less distorted 
(effective $g_0$ decreases),
resistivity decreases, and band like motion persists longer (i.e.,
the peak position of the resistivity shifts to higher temperatures).
Thus the system can have a large drop in resistivity at $T_C$
when a magnetic field is applied.

\section{ADIABATIC TRANSPORT OF SMALL POLARONS}
To study transport we use double exchange modification
 to include effects due to  on-site Hund's
coupling between itinerant holes and localized electrons
and take the total hamiltonian to be
 \begin{eqnarray} 
H=
- t_{DE}
\sum _{\langle i j \rangle} 
  c _{i }^{\dagger}  
  c _{j } 
  && 
+\sum_{\vec{q}} 
\omega_{\vec{q}}
a_{\vec{q}}^{\dagger}
a_{\vec{q}} 
 \nonumber \\                           
  && 
+\frac{1}{N^{1/2}} \sum_{j, \vec{q}}
  c _{j }^{\dagger} 
  c _{j } 
e^{i \vec{q} \cdot  \vec{R}_{j}}
g_{\vec{q}}
\omega_{\vec{q}}
( a_{\vec{q}} +
a_{-\vec{q}}^{\dagger}) ,
\label{HTDE}
 \end{eqnarray} 
where
$  c _{j } ~ (a_{\vec{q}})$ is the
 hole (phonon) destruction operator,
$ \langle i j \rangle$ corresponds to nearest neighbors,
$\omega_{\vec{q}}$ is the optical phonon frequency ($\hbar = 1$),
 $g_{q}\omega_{\vec{q}}$ is the hole-phonon coupling, and
$ t_{DE}= t \sqrt{(1+M^2/M_{S}^2)/2}$,
 $M$ the magnetization, $M_{S}$ the saturated
magnetization, and $N$ is the number of sites.
 Here it should be mentioned that by allowing
only one electron per site the restriction on two electrons
of opposite spin to occupy the same site is equivalent
to Pauli blocking and
can be more severe than hard core repulsion
 (i.e., it can lead to higher energies).
 Furthermore the above hamiltonian corresponds to assuming
a single orbital per site which on account of Jahn-Teller
splitting may 
 be justified.
Actually, we feel that only one orbital ought to be involved
in the transport process as tunneling between two
similar potential wells is more likely than between
two dissimilar potential wells (i.e., tunneling between
same orbitals is more likely because of resonant
tunneling). Then the electron lowers its energy better
through enhanced hopping or lower kinetic energy.
 Furthermore the experiments of A. Lanzara et al. \cite{lanzara}
are in agreement with our claims. We will elaborate on this
after we present our polaronic picture.

We will first outline the procedure for obtaining the resistivity
for the case where the hopping term is small compared to the binding
energy (i.e., the true small polaron case) and then proceed to
incorporate finite bandwidth effects. We will now perform
the so-called Lang-Firsov transformation
\cite{lang-firsov}
 $ \tilde{H} = e^S H e^{-S}$
to diagonalize the hamiltonian with
\begin{equation}
S = 
- \sum_{j, \vec{q}}
  c _{j }^{\dagger} 
  c _{j } 
e^{i \vec{q} \cdot  \vec{R}_{j}}
\frac{g_{\vec{q}}}{N^{1/2}}
( a_{\vec{q}} 
- a_{-\vec{q}}^{\dagger}) .
\end{equation}
The resulting hamiltonian is given by
 \begin{eqnarray} 
\tilde{H}=
- t_{DE}
\sum _{\langle i j \rangle} 
  c _{i }^{\dagger}  c _{j } 
  X _{i }^{\dagger}  X _{j } 
  && 
+\sum_{\vec{q}} \omega_{\vec{q}}
a_{\vec{q}}^{\dagger}
a_{\vec{q}} 
 \nonumber \\                           
  && 
- \sum_{j, \vec{q}} 
\frac {g_{\vec{q}}^2 \omega_{\vec{q}}}{N}
  c _{j }^{\dagger} c _{j } ,
\label{Htilde}
 \end{eqnarray} 
where
\begin{equation}
X_{j} = 
\exp \left [ \sum_{\vec{q}}
e^{i \vec{q} \cdot  \vec{R}_{j}}
\frac{g_{\vec{q}}}{N^{1/2}}
( a_{\vec{q}} 
- a_{-\vec{q}}^{\dagger}) \right ] ,
\label{Xj}
\end{equation}
and $\sum_{\vec{q}} \frac {g_{\vec{q}}^2 \omega_{\vec{q}}}{N}$ is
the binding energy.
Here it should be pointed out that to obtain the above transformed
hamiltonian we have used the following approximation 
 \begin{equation} 
 \sum_{j, k, \vec{q}} 
\frac {g_{\vec{q}}^2 \omega_{\vec{q}}}{N}
  c _{j }^{\dagger} c _{j } 
e^{i \vec{q} \cdot ( \vec{R}_{j}
- \vec{R}_{k} )}
  c _{k }^{\dagger} c _{k } 
= \sum_{j, \vec{q}} 
\frac {g_{\vec{q}}^2 \omega_{\vec{q}}}{N}
  c _{j }^{\dagger} c _{j } .
\label{approx}
 \end{equation} 
The above relationship is exact
when $ g_{\vec{q}}^2 \omega_{\vec{q}}$ is independent of $\vec{q}$.

The single small polaron eigen state of the hamiltonian in
Eq.\ (\ref{HTDE}) is given by
 \begin{equation} 
|\psi _{i} \rangle =
 e^{-S}|i \rangle |..n_{\vec{q}}.. \rangle 
= |i \rangle X_i |..n_{\vec{q}}.. \rangle ,
\label{psi_i}
 \end{equation} 
where $|i \rangle$ is the molecular orbital eigenstate at site $i$ and
$ |..n_{\vec{q}}.. \rangle $ is the product of the eigenstates
of the molecules at various sites executing simple harmonic motion
with phononic occupation number $ n_{\vec{q}}$.
The above wavefunction is exact in the limit of 
the ratio of the hopping term to the binding energy
being vanishingly small.

In Eq.\ (\ref{Htilde}) the first term involving the hopping term is
the small parameter. If the dominant transport mechanism corresponds
to diagonal processes (i.e., number of phonons in each state $\vec{q}$
remains unchanged) phase coherence is maintained when
the particle propagates. In fact then the particle moves like
a Bloch  electron and forms energy bands with  energy being
given by the thermal average of the first term 
in Eq.\ (\ref{Htilde})
 (see Appendix A for details).
 \begin{eqnarray} 
-\langle t_{DE}
\sum _{\langle i j \rangle} 
  c _{i }^{\dagger}  c _{j } 
  X _{i }^{\dagger}  X _{j } \rangle
&&=- 2 t_{DE} \sum _{\vec{k}} ( \cos (k_x a) + \cos (k_y a) + ..)
n_{\vec{k}}
  \langle X _{i }^{\dagger}  X _{j } \rangle
\nonumber \\
&& =
 - 2 t_{DE} \sum _{\vec{k}}( \cos (k_x a) + \cos (k_y a) + ..)
n_{\vec{k}}
\exp [-g_0^2 \coth( \beta \omega _0/2 ) ] 
\nonumber \\
&& =
 \sum _{\vec{k}} \epsilon _{\vec{k}}
n_{\vec{k}}
\label{Ek}
 \end{eqnarray} 
where 
$n_{\vec{k}}
= \langle c^{\dagger}_{\vec{k}} c_{\vec{k}} \rangle$ and
$\langle ... \rangle $ corresponds to thermal average.

We will now calculate the conductivity for localized states,
i.e., the hopping conductivity. The polarization operator
is given by $\vec{P} = \sum_{i} \vec{R}_{i} 
  c _{i }^{\dagger}  c _{i } $.
Then the current operator is given by
 \begin{eqnarray} 
 \vec{j} = q_e \frac{\partial \vec{P}}{\partial t} 
&&  =\frac{q_e i}{\hbar} \left [ \tilde{H} , \vec{P} \right ]
\nonumber \\
&& = - \frac{i q_e t_{DE}}{\hbar}
\sum _{\langle i j \rangle} 
[\vec{R}_{i} -
\vec{R}_{j}] 
  c _{i }^{\dagger}  c _{j } 
  X _{i }^{\dagger}  X _{j } .
 \end{eqnarray} 
Using the above form of the current operator and the 
many-body states obtained from the single particle states
$ |i \rangle |..n_{\vec{q}}.. \rangle $
 and taking $\omega _{\vec{q}} = \omega _0$ and $g_{\vec{q}} = g_0$,
we can obtain the conductivity to be (see Appendix B)
 \begin{eqnarray} 
{\rm Re}(\sigma _{\alpha \alpha}) && = 
\frac{1 - e ^{-\beta \omega }}{2 \omega}
\int _{-\infty}^{\infty} d t e^{i \omega t}
\langle j_{\alpha} (t) j_{\alpha} (0) \rangle
\nonumber \\
&& = \frac{n_c q_{e}^2 a^2}{k_B T} 
\frac{\sqrt{\pi}t_{DE}^2}{g_0 \omega _0 
\sqrt{{\rm cosech \omega_0 \beta /2}}}
e^{-2 g_0^2 
{\rm tanh}(\beta \omega_0 /4)} ,
\label{sigma}
 \end{eqnarray} 
where $n_c$ is the density of carriers.
 Furthermore, it should be noted 
that we need $2 g_0 ^2
{\rm csch}(\beta \omega_0/2) >> 1$
for Eq.\ (\ref{sigma}) to be valid.
Now the mobility, for a system of non-degenerate electrons,
 is given by the Einstein relation $\mu = q_e D \beta$
where 
$D$ is the diffusivity.
 In the region of interest, i.e., around
the metal-insulator 
transition, we expect band narrowing to be sufficiently
 strong so that the Fermi energy 
is not much larger than the thermal energy. Since the hopping-regime
 diffusivity is given by $D_{hop}=a^2/(6 \tau )$, we readily 
obtain the scattering time $\tau$ to be
 \begin{equation} 
\frac{1}{6 \tau}= 
\frac{\sqrt{\pi} t_{DE}^2}{g_0 \omega _0
 \sqrt{{\rm cosech \omega_0 \beta /2}}}
e^{-2 g_0^2 
{\rm tanh}(\beta \omega_0 /4)} .
 \end{equation} 
The above expression for $\tau$ corresponds to the non-adiabatic
regime (or $ t << \omega_0$). As for the adiabatic case, $\tau$ is 
given by
 \begin{equation} 
\frac{1}{6 \tau}= 
\frac{\omega_0}{2 \pi}
e^{-2 g_0^2 
{\rm tanh}(\beta \omega_0 /4)} ,
\label{tau1}
 \end{equation} 
which for high temperatures reduces to the classical case
 \begin{equation} 
\frac{1}{6 \tau}= 
\frac{\omega_0}{2 \pi}
e^{- g_0^2 \omega _0
\beta /2 } ,
\label{tau2}
 \end{equation} 
with $ g_0^2 \omega_0 /2$ corresponding to the activation energy.

The diffusivity for band conduction is given by
 \begin{equation} 
D_{band}= \langle |\vec{\nabla} E_{\vec{k}}|^2
 \tau \rangle/{ d} =2 \tau a^2 t^{2}_{DE} 
e^{-2 g_0^2 
{\rm coth}(\beta \omega_0 /2)} ,
\label{Dband}
 \end{equation} 
where $d$ is the dimension of the system and $\tau$ is given
by the same expression as in the hopping case.\cite{sudhakar1}

Then based on Friedman's work \cite{friedman} we take
the total mobility 
($\mu _T$)
to be the  sum of the band mobility and the hopping mobility and
hence the total resistivity ($1/\rho = n_c q_e \mu_T$) to be given by
 \begin{eqnarray} 
\frac{4 \pi}{
 n_c q_{e}^{2} a^2 \rho} =
 \beta \omega_0  && \left [  \frac{8 \pi ^2} {3}  
\frac{t^2}{\omega_{0} ^2}
 \exp {[-2 \theta 
 {\rm cosech} (\frac{\beta \omega_0}{2})]} + \right . 
 \nonumber \\                           
  && 
\left . \left ( 1+  \frac{M^2}{M_{S}^2} 
 \right ) \exp{[-2 \theta \tanh 
(\frac{\beta \omega_0}{4})]} \right ],
\label{rho_t=0}
 \end{eqnarray} 
where 
$\theta =g_0 ^2$.
Here we note that Friedman's analysis also accounts for 
how an electron may seem localized yet it can have translation
invariance symmetry intact -- the reason being
that the mobility is the sum
of the hopping and band-like mobilities and hence will always have
a component, no matter how small, that is metallic.

\section{SMALL POLARON PICTURE WITH FINITE BANDWIDTH EFFECTS}
We will now include finite bandwidth effects in our small
polaronic picture. However we will
still treat the ratio 
$\frac{t_{DE}}{2 g_0^2  \omega_0}$ 
as small. The polaronic wave function now
spreads and barely extends to the nearest neighboring sites.
Then the eigenstate of the nearly small
polaron is given by (see Toyozawa's paper also \cite{toyozawa})
 \begin{equation} 
|\Psi _{i} \rangle =
 \sum _{j} B(j)|i-j \rangle \tilde{X}_i
 |..n_{\vec{q}} .. \rangle ,
\label{Psi_i}
 \end{equation} 
where $\tilde{X}_i$ is given by
\begin{equation}
\tilde {X}_i = 
\exp \left [
 \sum_{ \vec{q}}
e^{i \vec{q} \cdot  \vec{R}_{i}}
\alpha _{\vec{q}}
( a_{\vec{q}} 
- a_{-\vec{q}}^{\dagger}) \right ] .
\end{equation}
In the above equation
 $B(j) =0$ for $|\vec{R}_j| > a$.
 Furthermore $B(j)$ and $\alpha_{\vec{q}}$ are
to be determined by minimizing the
single small polaronic energy.
Upon taking the expectation value of the hamiltonian
with respect to a small polaron state of momentum $\vec{k}$ given by
$|\Psi _{\vec{k}} \rangle = \sum_{i} e^{\vec{k} \cdot \vec{R}_{i}}
|\Psi _{i} \rangle $
 one gets
\begin{equation}
-\langle \Psi _{\vec{k}} |
 t_{DE}
\sum _{\langle i j \rangle} 
  c _{i }^{\dagger}  
  c _{j } 
|\Psi _{\vec{k}} \rangle \approx  -
t_{DE} \sum_{\delta , i} B(i) B(i - \delta )
\end{equation}
and
\begin{eqnarray}
\langle \Psi _{\vec{k}} | &&
\sum_{\vec{q}} 
\omega_{\vec{q}}
a_{\vec{q}}^{\dagger}
a_{\vec{q}} 
+\frac{1}{N^{1/2}} \sum_{j, \vec{q}}
  c _{j }^{\dagger} 
  c _{j } 
e^{i \vec{q} \cdot  \vec{R}_{j}}
g_{\vec{q}}
\omega_{\vec{q}}
( a_{\vec{q}} +
a_{-\vec{q}}^{\dagger}) 
|\Psi _{\vec{k}} \rangle
\nonumber \\
 &&
 \approx \sum _{\vec{q}} \omega_{\vec{q}}
( N_ {\vec{q}} + \alpha _{\vec{q}}^2 )
- 2 N^{-1/2} \sum _{\vec{q} , i} 
\alpha_{\vec{q}}
\omega _{\vec{q}} g_{\vec{q}}
\cos (\vec{q} \cdot \vec{R}_{i} ) B^2 (i) ,
\end{eqnarray}
where we have neglected the small valued vibrational overlap factors
$\langle .. n_{\vec{q}}^{\prime}  ..|
\tilde{X}^{\dagger}_{i} \tilde{X}_{j}|.. n_{\vec{q}}  .. \rangle $ 
 for $i \neq j$.
Upon minimizing the polaron energy with respect to $\alpha_{\vec{q}}$
one obtains
\begin{equation}
\alpha _{\vec{q}}
 =
N ^{-1/2} g_{\vec{q}} \left [ \sum _{j} B(j)^2
\cos ( \vec{q} \cdot  \vec{R}_{j}) \right ] .
\end{equation}
 For small values of $\frac{t_{DE}}{2 g_0^2  \omega_0}$,
on using the constraint that $\sum_{j} B^2 (j) = 1$,
 one obtains
from the above equations
$B(0) \approx 1$ and
\begin{equation}
B(j) \approx \frac{t_{DE}}{2 N^{-1} \sum_{\vec{q}} \omega_{\vec{q}} 
g_{\vec{q}}^2  
(1 - \cos ( \vec{q} \cdot  \vec{R}_{j}))} , 
\end{equation}
 for $|\vec{R}_{j}| = a$.
The above results are similar to those obtained by Gosar.\cite{gosar}

Using the above eigenstate 
$|\Psi _{i} \rangle$ for the small polaronic state and again using
the procedure outlined in Appendix B for deriving the conductivity
in the localized regime, 
one obtains (on neglecting $B(j)$ for $j \neq 0$ due to its 
small contribution) the same expression for conductivity
but with the lattice distortion $g_0^2$ renormalized
\begin{equation}
{\rm Re}[\sigma _{\alpha \alpha}(\omega )]
 = \frac{1 - e ^{-\beta \omega }}{6 \omega}
 \frac{ q_{e}^2 t_{DE}^2 }{\hbar ^2} c(1-c) a^2 \frac{z N}{V}
\sqrt{\pi / \tilde{\gamma}} e^{-2 \tilde{S}_T +
 \tilde{\phi} (-i\beta /2 )+
 \omega \beta /2 
 -\omega ^2 /(4 \tilde{\gamma}) } ,
\end{equation}
where 
\begin{equation}
\tilde{S}_T
\equiv
\sum_{\vec{q}} 
\frac{ |\tilde{\lambda} _{\vec{q}}|^2}{2}
 \coth \frac{\beta \omega_{\vec{q}}}{2} ,
\end{equation}
\begin{equation}
\tilde{\phi}(t) = \sum _{\vec{q}} |\tilde{\lambda} _{\vec{q}}|^2
[ (N_{\vec{q}} + 1)  e^{-i \omega_{\vec{q}}t} +
N_{\vec{q}} e^{i \omega_{\vec{q}}t}] ,
\end{equation}
and 
\begin{equation}
\tilde{\gamma}
=\sum_{\vec{q}}
|\tilde{\lambda} _{\vec{q}}|^2 
[ N_{\vec{q}}
(N_{\vec{q}} + 1) ]^{1/2}
 \omega_{\vec{q}}^2 ,
\end{equation}
with
$\tilde{\lambda} _{\vec{q}} = e^{{ i\vec{q}}\cdot \vec{R}_{j}}
 \left ( 1- e^{i \vec{q}
\cdot \vec{\delta}} \right ) \alpha_{\vec{q}} $.
Here it should be pointed out that the authors of Ref.\ \onlinecite{lee}
did not take into account the renormalization
of $\lambda _{\vec{q}}$ due to the finite bandwidth effects.
They also failed to recognize that $ \tilde{\gamma} \beta >>
 |\epsilon_{\vec{k}} - \epsilon _{\vec{p}}| $ 
while
evaluating the integral with respect to time
in order to obtain the expression for conductivity.

Now, upon 
taking $\omega _{\vec{q}} = \omega _0$ and $g_{\vec{q}} = g_0$,
 one obtains the same expression for the
total resistivity given by Eq.\ (\ref{rho_t=0}) where  $\theta$ is
expressed as follows
\begin{equation}
\theta \equiv g_0 ^2 \left [
1 - \frac{(z+1) t_{DE} ^2}{2 g_0 ^4 \omega_{0}^2 } \right ] ,
\label{theta}
\end{equation}
with
$z$ being the coordination number.\cite{sudhakar2}
 Here, it should be noted 
that we need $2 \theta {\rm csch}(\beta \omega_0/2) >> 1$
for Eq.\ (\ref{rho_t=0}) to be valid.
 Furthermore, the optical conductivity above the metal-insulator 
transition (where only the conductivity due to localized 
carriers dominates) is given by
\begin{equation}
\frac{{\rm Re}[\sigma _{\alpha \alpha}(\omega )]}
{\sigma _{\alpha \alpha}(\omega \rightarrow 0 )}
 = \frac{\sinh(\beta \omega /2 )}{\beta \omega /2}
 e^{ -\omega ^2 /(4 \tilde{\gamma _0}) } ,
\label{optcond}
\end{equation}
where $\tilde{\gamma _0} =\theta \omega _0 ^2
{\rm cosech}(\beta \omega _0 /2 )$.
We note, from Eq.\ (\ref{optcond}), that the  optical conductivity 
 scaled by the
DC conductivity depends only on the
parameter $\tilde{\gamma _0}$.

We will now study the magnetic transition
 within a mean field approximation.
The magnetization ratio $M/M_S$
is not very sensitive  to the type of
approximation (see Ref.\ \onlinecite{sudhakar3} for
 a weak Hund's coupling treatment
of the magnetization problem).
The magnetization ratio is given by
 \begin{equation} 
S\frac{M}{M_{S}}=
- \frac {\sum_{S_z} S_{z} \exp[ -
 g  \mu_{B} H_{eff} 
 S_z \beta]}
{\sum_{S_z} \exp[ -
g  \mu_{B} H_{eff} 
S_z \beta]} ,
\label{magratio}
 \end{equation} 
with $H_{eff} =\lambda \frac{M}{M_S}$.
On using the condition that as $T \rightarrow T_C$ we have
$M/M_S \rightarrow 0$, we get for $S=3/2$ the relation
 $\lambda g \mu_{B}
= 1.2 k_{B} T_C$.
Estimating the transition temperature $T_C$ with accuracy is difficult
and we will only give an order of magnitude estimate for it and
will provide a qualitative feel for its dependence on various
Physical parameters of the system.
Above the transition point the electrons are localized
and form small ferromagnetic domains (or magnetic polarons)
so as to minimize the free energy.
At the transition point the magnetic polarons align to give
a ferromagnetic phase whose size is of the order of the
size of the system. At this point the decrease in kinetic energy because
of the electronic delocalization is equal to the increase in the
entropic contribution to the free energy as given below.\cite{Meq1}
 \begin{equation} 
- N \delta 6 t_{DE} e ^{- \theta 
{\rm coth}(\beta \omega_0 /2) }
 \approx -N 
 k_{B} T_C \ln 4 .
 \end{equation} 

\section{RESULTS AND DISCUSSION}

In the doping regime where the manganites are insulating
(i.e., $\delta \sim 0 $ or $\delta > 0.5$),
although there is orbital order and
both orbitals enter the hamiltonian, in the doping regime
 $ 0.2 < \delta < 0.4$ where CMR is observed
only one orbital need be considered. As described
in a recent interesting paper by Khomskii \cite{khomskii},
the manganite system tries to lower its overall energy by
entering into a ferromagnetic orbitally ordered state
with the same orbital being occupied at each site.
The situation is similar to what is encountered in Nagaoka
type of ferromagnetism in spin systems. When doped with a few
holes, just as an antiferromagnetic spin state can become ferromagnetic
so will an orbitally aniferromagnetic state become an orbitally
ferromagnetic one. In Ref.\ \onlinecite{khomskii},
it is also pointed out that at sizeable doping a 
state with $d_{x^2 - y^2}$
or $d_{z^2}$ ordering may have lower energy than the
proposed state where only 
 $d_{z^2} \pm i d_{x^2 - y^2}$ orbitals are occupied.

Based on the experimental results of Lanzara et al. \cite{lanzara} 
we will now try to justify that only one of 
the orbitals $d_{z^2}$ or $ d_{x^2 - y^2}$ is occupied and that the
orbital ordering temperature is higher than the magnetic transition
temperature $T_C$ (which is possible because there is no reason
to expect coupling between
the order parameters for the magnetic transition and the orbital
ordering transition).
In Ref.\ \onlinecite{lanzara}, in Fig. 4 we see that only
one type of distortion of the
 octahedron (the so-called $Q_3$ normal mode)
seems to be relevant both above and below $T_C$. However,
there are two distortions of this same type (at sites A and B)
 above $T_C$ but their degree of distortion is different.
We think that it indicates that the lattice distortion
is less in the ferromagnetic domains (site A)
that exist even above $T_C$
and is similar to  the distortion (again of the same $Q_3$ type)
 in the ferromagnetic
region below $T_C$ at $T < 200~K$. 
Moreover, in the paramagnetic regime (site B) the
distortion is expected to be more in our
picture because
the effective hopping integral
$t_{DE}$ is smaller in this regime. 
 Furthermore, from Fig. 2 of Ref.\ \onlinecite{lanzara} we see
that even at $T = 300~K$ ($>  T_C = 240~K$) only $Q_3$ mode exists
which prompts
us to conclude that the orbital ordering probably occurs at a 
fairly higher temperature than $T_C$.
 It is also of interest to note from Fig. 3 of 
Ref.\ \onlinecite{lanzara} that
EXAFS probes instantaneous and local distortions
that are larger than the ones observed from diffraction experiments.
This may explain why other probes (like neutron scattering)
do not show noticeable JT distortion at low temperatures.

Our magnetization curves $M/M_S$ as a function of the
reduced temperature $T/T_C$ [see Fig. \ref{fig1}]
are independent of the values of the various parameters
of the system like $t$, $\delta$, $\omega_0$, and $g_0$ because
of the mean-field nature of the approximation.
The qualitative behavior of the experimental curves is
mimicked by our calculations but the experimental
values of $M/M_S$ rise faster with $T/T_C$ 
(see Ref.\ \onlinecite{urushibara}).

The peak in the resistivity occurs when the system goes from
insulating behavior to a metallic behavior (however the peak need
not occur exactly when the hopping mobility becomes equal to
the band mobility). When the system becomes metallic the system also
becomes ferromagnetic because the itinerant electrons polarize the 
localized spins. Thus we can take the metal-insulator transition
point as also the magnetic transition point -- a fact borne out
by experiments (see Ref.\ \onlinecite{urushibara}).

 From the expression for the resistivity (see Eq.\ (\ref{rho_t=0}))
it follows that for a given value of $t/\omega_0$, the ratio
$k_B T_C/ \omega_0$ is fixed and one need not
 treat $\omega_0$ as a variable
when studying resistance dependence  on various parameters.

We will now discuss the resistivity 
given by Eq.\ (\ref{rho_t=0}).
 The conduction goes from a hopping 
type at high temperatures to a band 
 type at low temperatures.
In Fig.\ \ref{fig2} we have shown 
 the dependence of resistivity $\rho$ on temperature
at various magnetic fields. 
The values of the hopping integral $t$ are taken 
such that the bandwidth lies in the range $1~eV - 3.5~eV$
which is a realistic range based on band structure calculations.
The values of $g^2_0$ are taken from the experimentally
obtainable activation energy ($\theta \omega_0 /2$)
values corresponding to temperatures in the range $1000~K - 2500~K$
(see Refs.\ \onlinecite{ramirez,worledge}) while the chosen value
of the Debye temperature $ T_D = 500~K$
is realistic too (see Ref.\ \onlinecite{ramirez}).
The general trend of the resistivity including the drop
at the MI transition at $H = 0~T$
 is similar to the experimental results.\cite{urushibara}
On introducing a magnetic
field the system gets magnetized at temperatures higher than $T_{C}$
and thus the value of $\theta$ is smaller
 (see Eq.\ (\ref{theta})).
 Consequently the resistivity is smaller and 
 $T_{\rho^{max}}$ (the temperature at which resistivity
becomes maximum) increases.

 For $T \geq T_C$, 
when  $D_{band}/D_{hop} >>1$  
 the magnetoresistance 
$\Delta \rho (H) \equiv (\rho(0)-\rho(H))/\rho(0)$
 is given by [see Eqs.\ (\ref{rho_t=0}) and (\ref{theta})]
 \begin{equation} 
\Delta \rho(H) 
\approx 
1 - \exp \left [ - \frac {(z+1)}{2 g_0 ^2} \frac{ t^2}{\omega_{0} ^2} 
\frac{ M^2}{ M_{S}^2}
{\rm csch}(\frac{\beta \omega_{0}}{2} ) \right ] ,
\label{deltarhoband}
 \end{equation} 
and when $D_{band}/D_{hop} <<1$ it is given by
 \begin{equation} 
\Delta \rho(H) 
\approx 1 - \frac{\exp 
\left [ -\frac{(z+1)} {2 g_0^2 }
\frac{t^2}{ \omega_{0}^2}
 \frac{ M^2} { M_{S}^2} 
 \tanh(\frac{\beta \omega_{0}}{4} ) \right ] 
 }{1 + (M/M_{S})^2} .
\label{deltarhohop}
 \end{equation} 

 For $\beta \omega_{0}/2 < 1$, on taking
${\rm csch}(\beta \omega_0/2) \approx 
2/(\beta \omega_{0})$  and 
$\tanh (\beta \omega_0/4) \approx 
\beta \omega_{0}/4$, 
if $\theta \beta \omega_{0}/2 > 1$ 
the following can be shown: 
(i) $T_{\rho_{M}^{max}}$ increases as $g_0 ^2$ decreases
(or $t^2/\omega_{0}^2$ increases); and (ii) for fixed values of
$g_0^2 $ and $t^2/\omega_{0}^2$
 and for large enough
 $\frac{(z+1)}{2 g_0^2} \frac{t^2}{\omega_{0}^2}$,
 $T_{\rho^{max}}$ increases as $M$ increases.
The above observations are borne out by the numerical results
reported in Table I where the
 empty boxes 
correspond to cases where our approximation may not be good.
 We further note that for the same value of
the polaron size parameter
$t/(\omega_0 g_0^2)$ the magnetoresistance $\Delta \rho (H)$
increases as the adiabaticity parameter 
$t/{\omega_0}$ increases.

In Fig.\ \ref{fig3}, we plot the scaled optical conductivity
${\rm Re} \sigma (\omega ) / \sigma (0) $
[given by Eq.\ (\ref{optcond})] as a function
of the frequency at different temperatures.
The maximum of the optical conductivity occurs at
$\omega \approx \tilde{\gamma _0} \beta$ as expected from the formula
in Eq.\ (\ref{optcond}). We note that
as the value of the renormalized electron-phonon coupling
parameter $\theta$
increases, the optical conductivity curve spreads out more.
We also find that, as the temperature increases the value of the scaled
optical conductivity decreases.
 Furthermore, the  calculated scaled curves are in qualitative 
agreement with experiments. In the experimental
situation there are two pieces to the conductivity --
one coming from transitions with electrons parallel to the core
spins ($\omega \sim t$) and another at higher
energy ($\omega \sim $ twice the Hund's coupling energy) involving
transitions to states
where the electron spins are antiparallel
to the core spins. However, since we do not allow
for double occupancy at any site,
the second piece of the optical conductivity
does not appear in our calculated curves.

In conclusion, we say that we showed the importance of finite
bandwidth effects in understanding CMR within a small polaron
picture. In addition 
to the polaron size parameter studied by other authors \cite{millis2},
we have also identified another dimensionless
parameter (the adiabaticity parameter) and demonstrated its importance.
The values of magnetoresistance calculated by us compare favorably
with the experimentally reported ones.

\begin{center}
{\bf ACKNOWLEDGMENTS}
\end{center}

{The author would like to thank C. S. Ting, T. V. Ramakrishnan, 
H. R. Krishnamurthy, V. Pai, D. D. Sarma, Ram Seshadri, 
Jinwu Ye, G. F. Giuliani, 
and A. N. Das for discussions. The author would also like to acknowledge
useful discussions with D. Khomskii regarding the validity of 
considering only one orbital in the doping regime where CMR occurs.
This work was partially supported
by Texas Center for Superconductivity and a grant from
Texas ARP (ARP-003652-0241-1999).}

\appendix
\section{}
In this Appendix we will derive the band narrowing due to phonons.
The number of phonons in each state $\vec{q}$ remains unchanged in a
diagonal transition. If this is the dominant mechanism, then
phase coherence is maintained and the electron propagates
as a band-like particle. Using the decoupling scheme
  $\langle c _{i }^{\dagger}  c _{j } 
  X _{i }^{\dagger}  X _{j } \rangle
  = \langle c _{i }^{\dagger}  c _{j } \rangle \langle 
  X _{i }^{\dagger}  X _{j } \rangle$
 one obtains
the single particle energy to be
\begin{eqnarray}
\epsilon_{\vec{k}} && =-2 t_{DE} ( \cos (k_x a) + \cos (k_y a) + ..)
  \langle X _{j+\delta}^{\dagger}  X _{j } \rangle
\nonumber \\
&& = - 2 t_{DE} ( \cos (k_x a) + \cos (k_y a) + ..)
\exp [-g_0^2 \coth( \beta \omega _0/2 ) ] ,
\label{app1}
\end{eqnarray}
where $|\vec{R}_{\delta}| = a$. We will now proceed to derive the
above expression.
Now
\begin{equation}
  \langle X _{j+\delta}^{\dagger}  X _{j } \rangle=\Pi _{\vec{q}}
\langle e^{\lambda _{\vec{q}} a _{\vec{q}}
- \lambda _{\vec{q}}^{\star} a _{\vec{q}}^{\dagger}} \rangle ,
\label{app2}
\end{equation}
where
$\lambda _{\vec{q}} = e^{{ i\vec{q}}\cdot \vec{R}_{j}}
 \left ( 1- e^{i \vec{q}
\cdot \vec{\delta}} \right ) \frac{g_{\vec{q}}}{N^{1/2}} $.
Then the thermal average is given by
\begin{equation}
  \langle X _{j+\delta}^{\dagger}  X _{j } \rangle
= \Pi _{\vec{q}} 
\frac 
{ \sum _{n_{\vec{q}} =0 }^{\infty}
\langle n_{\vec{q}} |
 e^{-\beta n_{\vec{q}}
\omega _{\vec{q}}}
 e^{-|\lambda _{\vec{q}}|^2 /2}
 e^{ - \lambda _{\vec{q}}^{\star} a _{\vec{q}}^{\dagger}} 
 e^{\lambda _{\vec{q}} a _{\vec{q}}}
  | n_{\vec{q}} \rangle }
{ \sum _{n_{\vec{q}} =0 }^{\infty}
\langle n_{\vec{q}} |
 e^{-\beta n_{\vec{q}}
\omega _{\vec{q}}}
  | n_{\vec{q}} \rangle } .
\label{app3}
\end{equation}
On noting that 
\begin{equation}
e^{u a } |n \rangle = \sum _{l = 0} ^{\infty}
\frac{u ^l}{l !} a ^l | n \rangle ,
\label{app4}
\end{equation}
with
\begin{equation}
 a ^l | n \rangle  = \left [ \frac{n !}{(n-l) !} \right ] ^{1/2}
 |n - l \rangle ,
\label{app5}
\end{equation}
we get the following relationship
\begin{equation}
\langle n |
 e^{- u ^{\star} a^{\dagger} }
 e^{ u a }
 |n \rangle = \sum _{l = 0} ^{n}
\frac{ \left ( -|u|^2 \right ) ^{l}}{(l !)^2} 
 \left [ \frac{n !}{(n-l) !} \right ] = L_n (|u|^2) ,
\label{app6}
\end{equation}
where $L_n (x)$ is the Laguerre polynomial.
Since the following identity holds
\begin{equation}
\sum _{l = 0} ^{\infty}
 L_n (|u|^2) z^l = \frac{e^{\left [ |u|^2 \frac{z}{z-1} \right ] }}{1-z},
\label{app7}
\end{equation}
we obtain from Eqs.\ (\ref{app3}) -- (\ref{app7})
\begin{equation}
  \langle X _{j+\delta}^{\dagger}  X _{j } \rangle
= \Pi _{\vec{q}}
 e^{-|\lambda _{\vec{q}}|^2 /2}
 e^{-|\lambda _{\vec{q}}|^2 N_{\vec{q}}}
=e^{- \sum_{\vec{q}} 
\frac{ |\lambda _{\vec{q}}|^2}{2} \coth 
\frac{\beta \omega_{\vec{q}}}{2}} 
\equiv e^{-S_T} ,
\label{app8}
\end{equation}
with $N_{\vec{q}}$
being the Bose-Einstein distribution function.
Then for $\omega_{\vec{q}} = \omega_0$ and $g_{\vec{q}} = g_0$
we obtain Eq.\ (\ref{app1}).

\section{}

In this Appendix we will
calculate the conductivity within the small polaron
picture in the hopping regime.
 \begin{eqnarray} 
{\rm Re} && [\sigma _{\alpha \alpha} (\omega )]
\nonumber \\
 = && 
\frac{1 - e ^{-\beta \omega }}{2 \omega}
\int _{-\infty}^{\infty} d t e^{i \omega t}
\langle j_{\alpha} (t) j_{\alpha} (0) \rangle
\nonumber \\
 = &&
 \frac{ e^2 t_{DE}^2 }{\hbar ^2} 
\frac{1 - e ^{-\beta \omega }}{6 \omega}
\sum _{\delta , \delta ^{\prime} , j , j ^{\prime}} 
\vec{\delta} \cdot \delta ^{\prime}
\int _{-\infty}^{\infty} d t e^{i \omega t}
\langle 
  c _{j } ^{\dagger}(t)
  c _{j  + \delta} (t)
  c _{j^{\prime} + \delta ^{\prime} }^{\dagger}  c _{j ^{\prime} } 
\rangle
\langle 
X _{j } ^{\dagger} (t)  
  X _{j  + \delta} (t)  
  X _{j^{\prime} + \delta ^{\prime} }^{\dagger}  X _{j ^{\prime} } 
\rangle .
\label{sigmaapp}
 \end{eqnarray} 
In the above equation, the dominant contribution is obtained
when $j=j^{\prime}$ and $\delta = \delta ^{\prime}$.
The first correlation function in the above equation can be
 approximated by
\begin{equation}
\langle 
  c _{j } ^{\dagger}(t)
  c _{j  + \delta} (t)
  c _{j + \delta }^{\dagger}  c _{j } 
\rangle
 = \frac{1}{N^2} \sum _{\vec{k} , \vec{p} } f_{\vec{k}} ( 1 -
f_{\vec{p}} ) e^{i(\epsilon _{\vec{k}} - \epsilon _{\vec{p}} ) t } ,
\label{appb1}
\end{equation}
where $f_{\vec{p}}$ is the Fermi-Dirac distribution function.
 Now
 \begin{eqnarray} 
\langle 
X _{j } ^{\dagger} (t) && 
  X _{j  + \delta} (t)  
  X _{j + \delta }^{\dagger}  X _{j } 
\rangle  
\nonumber \\
&& = 
\frac{{\rm Tr} \left \{ e^{-\beta \tilde{H}}
e^{i \tilde{H} t}
 X _{j } ^{\dagger}   
  X _{j  + \delta}   
e^{-i \tilde{H} t}
  X _{j + \delta }^{\dagger}  X _{j } 
\right \}}
{{\rm Tr} \left \{ e^{-\beta \tilde{H}} \right \} }
\nonumber \\
 && =
\Pi _{\vec{q}}
\frac
 { \sum _{n_{\vec{q}} =0 }^{\infty}
\langle n_{\vec{q}} |
 e^{-\beta n_{\vec{q}}
\omega _{\vec{q}}}
 e^{-|\lambda _{\vec{q}}|^2 /2}
 e^{  \lambda _{\vec{q}}^{\star}
 a _{\vec{q}}^{\dagger} e^{i \omega _{\vec{q}}t }} 
 e^{- \lambda _{\vec{q}} a _{\vec{q}} e^{-i \omega_{\vec{q}}t}}
 e^{-|\lambda _{\vec{q}}|^2 /2}
 e^{ - \lambda _{\vec{q}}^{\star} a _{\vec{q}}^{\dagger}} 
 e^{\lambda _{\vec{q}} a _{\vec{q}}}
  | n_{\vec{q}} \rangle } 
 { \sum _{n_{\vec{q}} =0 }^{\infty}
\langle n_{\vec{q}} |
 e^{-\beta n_{\vec{q}}
\omega _{\vec{q}}}
  | n_{\vec{q}} \rangle } 
\nonumber \\
 && =
\Pi_{\vec{q}}( 1 - e^{-\beta \omega _{\vec{q}}})
 e^{-|\lambda _{\vec{q}}|^2 (1- e^{-i \omega_{\vec{q}}t})}
\sum_{n_{\vec{q}} =0}^{\infty}
 e^{-\beta n_{\vec{q}}
\omega _{\vec{q}}}
\langle n _{\vec{q}} | 
 e^{  \lambda _{\vec{q}}^{\star} a _{\vec{q}}^{\dagger} 
( e^{i \omega _{\vec{q}}t}- 1) } 
 e^{- \lambda _{\vec{q}} a_{\vec{q}} (e^{-i \omega_{\vec{q}}t } -1 )}
| n_{\vec{q}} \rangle  
\nonumber \\
 && =
\Pi_{\vec{q}}
 e^{-|\lambda _{\vec{q}}|^2[
(N_{\vec{q}} + 1) (1- e^{-i \omega_{\vec{q}}t })
 + N_{\vec{q}} (1- e^{i \omega_{\vec{q}}t})]} ,
\label{appb2}
 \end{eqnarray} 
where, to obtain the last line,
use has been made of the fact that
\begin{equation}
(1- e^{-\beta \omega _{\vec{q}}}) \sum _{n_{\vec{q}} =0}^{\infty}
 e^{-\beta n_{\vec{q}}
 \omega _{\vec{q}}}
\langle n_{\vec{q}} |
 e^{u^{\star} a^{\dagger}}
 e^{- u a}
|n_{\vec{q}} \rangle = e^{- |u|^2 N_{\vec{q}}} .
\label{appb3}
\end{equation}
Defining 
\begin{equation}
 \phi(t) = \sum _{\vec{q}} |\lambda _{\vec{q}}|^2
[ (N_{\vec{q}} + 1)  e^{-i \omega_{\vec{q}}t} +
N_{\vec{q}} e^{i \omega_{\vec{q}}t}] ,
\label{appb4}
\end{equation}
we have $\phi(t)= \sum _{\vec{q}} 
2 |\lambda _{\vec{q}}|^2 
[ N_{\vec{q}}
(N_{\vec{q}} + 1) ]^{1/2}
  \cos [\omega _{\vec{q}} (t + i\beta /2 )]$ and 
obtain from
Eq.\ (\ref{appb2})
 \begin{eqnarray} 
\int _{-\infty}^{\infty} dt e^{i \omega t} &&
\langle 
  c _{j } ^{\dagger}(t)
  c _{j  + \delta} (t)
  c _{j + \delta }^{\dagger}  c _{j } 
\rangle
\langle 
X _{j } ^{\dagger} (t)  
  X _{j  + \delta} (t)  
  X _{j + \delta }^{\dagger}  X _{j } 
\rangle  
\nonumber \\
 && =
\int _{-\infty}^{\infty} dt e^{i \omega t}
e^{-2 S_T} e^{\phi(t)}
\langle 
  c _{j } ^{\dagger}(t)
  c _{j  + \delta} (t)
  c _{j + \delta }^{\dagger}  c _{j } 
\rangle
\nonumber \\
 && =
 \frac{1}{N^2}
e^{-2 S_T} e^{\phi(-i \beta /2 )}
\int _{-\infty}^{\infty} dt e^{i \omega t}
e^{-
\gamma
 (t +i \beta /2 )^2}
 \sum _{\vec{k} , \vec{p} } f_{\vec{k}} ( 1 -
f_{\vec{p}} ) e^{i(\epsilon _{\vec{k}} - \epsilon _{\vec{p}} ) t } 
\nonumber \\
 && 
\approx
\sqrt{\pi / \gamma} c(1-c)
 e^{-2 S_T + \phi (-i\beta /2 )+ \omega \beta /2 
 -\omega ^2 /(4 \gamma) } ,
\label{appb5}
\end{eqnarray}
where $\gamma
=\sum_{\vec{q}}
|\lambda _{\vec{q}}|^2 
[ N_{\vec{q}}
(N_{\vec{q}} + 1) ]^{1/2}
 \omega_{\vec{q}}^2$
and $c$ is the number of carriers per unit site.
In evaluating the above integral we assumed that
$\sum_{\vec{q}}
|\lambda _{\vec{q}}|^2 
[ N_{\vec{q}}
(N_{\vec{q}} + 1) ]^{1/2} >> 1$
and used the saddle-point approximation.
 Furthermore, use has been made of the fact that
$  \gamma \beta >> 
| \epsilon _{\vec{k}} - \epsilon _{\vec{p}} |$. 
\cite{sudhakar4}

\begin{figure}
\caption{Plot of the magnetization ratio $M/M_{S}$ versus
the reduced temperature $T/T_{C}$
at magnetic fields 
 $H = 0~T$, $H = 15~T$, and $H = 30~T$.} 
\label{fig1}
\end{figure}

\begin{figure}
\caption{Plot of the resistivity $\rho$ in units of
 $4 \pi /(n_c q_{e}^2 a^2)$
versus temperature $T$
in 3 dimensions 
for adiabaticity parameter
 $t/ \omega_0 = 6$,
$g_0^2 =12$,
Debye temperature $T_D =500~K$, and for the following magnetic
fields:
(i)  $H = 0~T$; 
(ii) $H = 15~T$; 
and
(iii) $H = 30~T$.}
\label{fig2}
\end{figure}

\begin{figure}
\caption{Plot of the scaled optical conductivity
${\rm Re } \sigma (\omega ) / \sigma (0)$
as a function of frequency at various values of the renormalized
electron-phonon coupling parameter $\theta$ and for: (a) $T= 300~K$
($\approx T_C$) and
(b) $T= 500~K$. In Table I,
$\theta = 6.75$ corresponds to $t/\omega _0 = 6$ and $g_0^2 =12$;
$\theta = 5.89$ corresponds to $t/\omega _0 = 4$ and $g_0^2 =9$; and
$\theta = 4.83$ corresponds to $t/\omega _0 = 2$ and $g_0^2 =6$.}
\label{fig3}
\end{figure}

\begin{table}
\caption{Calculated values of 
 the transition temperature 
$T_C$ and
magnetoresistance 
$\Delta \rho(H)$ at $T_C$ 
 for $T_D =500~K$,
various values
of $t/ \omega_0$ and $g_0^2$, 
and magnetic fields 
$H=15~T$, 
 and $H=30~T$.}
\label{tab1}
\end{table}

\begin{tabular}{l|ccc|ccc|ccc}   \hline\hline
  & 
\multicolumn{3}{c|} {$g_0^2 =6$}&
\multicolumn{3}{c|} {$g_0^2 =9$}&
\multicolumn{3}{c} {$g_0^2 =12$}
  \\ \cline{2-10}    
$ t/ \omega_0$  &
$\Delta \rho ({\rm 15 ~T} )$ & ~ $\Delta \rho ({\rm 30 ~T})$ &~ $T_C $& 
$\Delta \rho ({\rm 15 ~T} )$ & ~ $\Delta \rho ({\rm 30~T})$ &~ $T_C $& 
$\Delta \rho ({\rm 15~T} )$ & ~ $\Delta \rho ({\rm 30~T})$ & ~ $T_C $
 \\ \hline
 ~2  &  37\%  & 49\% & 297~K & 34\% & 45\% & 245~K & 31\% & 41\% & 228~K
 \\ \hline
 ~3  &  46\%  & 62\% & 400~K & 48\% & 62\% & 264~K & 44\% & 57\% & 237~K
 \\ \hline
 ~4  &        &      &      & 62\% & 77\% & 295~K & 58\% & 72\% & 248~K
 \\ \hline
 ~6  &        &      &      &      &      &      & 81\% & 93\% & 290~K 
 \\ 
\hline\hline
\end{tabular}


\begin{references}
   \bibitem{ramirez}   
   A. P. Ramirez,
   J. Phys.: Condens. Matter {\bf 9},                      
   8171 (1997).

   \bibitem{bishop}   
   A. R. Bishop and H. R\"{o}der,
   Current Opinion Solid State Mater. Sci.
   {\bf 2},                      
   244 (1997).

   \bibitem{khomsawa}   
 D. I. Khomskii and G. A. Sawatsky,
   Solid State Commun. {\bf 102}, 87 (1997).               

   \bibitem{allen}   
 P. B. Allen and V. Perebeinos, Phys. Rev. B {\bf 60},
 10747 (1999).

   \bibitem{degennes}   
   P.-G. de Gennes,
   Phys. Rev. B  {\bf 118},                      
   141 (1960).                          

   \bibitem{furukawa}   
   N. Furukawa, J. Phys. Soc. Jpn. {\bf 63}, 3214 (1994); 
   {\bf 64}, 2734 (1995).

   \bibitem{millis1}   
   A. J. Millis, P. B. Littlewood, and B. I. Shraiman,
   Phys. Rev. Lett.  {\bf 74},                      
   5144 (1995).                           

   \bibitem{millis2}   
   A. J. Millis, B. I. Shraiman, and R. Mueller,
   Phys. Rev. Lett.  {\bf 77},                      
   175 (1996).                           

   \bibitem{roder}   
   H. R\"{o}der,  Jun Zang, and A. R. Bishop,
   Phys. Rev. Lett.  {\bf 76},                      
   1356 (1996).                           

   \bibitem{lee}   
   J. D. Lee and B. I. Min,
   Phys. Rev. B  {\bf 55},                      
   12454 (1997).                          

%

   \bibitem{jaime}   
 M. Jaime, M. B. Salamon, M. Rubinstein, R. E. Treece, J. S. Horowitz,
 and D. B. Chrisey,
   Phys. Rev. B  {\bf 54},                      
   11914 (1996).                          

   \bibitem{worledge}   
   D. C. Worledge, L. Mi\'{e}ville, and T. H. Geballe,
   Phys. Rev. B  {\bf 57},                      
   15267 (1998).                          

   \bibitem{ting}   
   We do not study localization due to nonmagnetic
 randomness. For a treatment of localization due
 to both spin disorder and nonmagnetic randomness
 see L. Sheng, 
 D. Y. Xing, D. N. Sheng, and C. S. Ting,
 Phys. Rev. Lett.  {\bf 79}, 1710 (1997); Phys. Rev. B
 {\bf 56}, R7053 (1997).
   
   \bibitem{toyozawa}   
   Y. Toyozawa, Prog. Theor. Phys. {\bf 26}, 29 (1961).
   


   \bibitem{urushibara}   
   A. Urushibara, Y. Moritomo, T. Arima, A. Asamitsu,
   G. Kido, Y. Tokura, 
   Phys. Rev. B  {\bf 51},                      
   14103 (1995).                          

   \bibitem{quijada}   
   M. Quijada, J. \v{C}erne, J. R. Simpson, H. D. Drew,
   K. H. Ahn, A. J. Millis, R. Shreekala, R. Ramesh, M. Rajeswari, and 
   T. Venkatesan, 
   Phys. Rev. B  {\bf 58},                      
   16093 (1998).                          

   \bibitem{hartman}   
   E. Muller-Hartman and E. Dagotto,
   Phys. Rev. B  {\bf 54},                      
   R6819 (1996).                          

   \bibitem{holstein}   
   T. Holstein,
  Annals of Physics
   {\bf 8},                      
   343 (1959).                          


   \bibitem{friedman}   
   L. Friedman, Phys. Rev. {\bf 135},                      
   A233 (1964).                          
 
  \bibitem{lanzara}
 A. Lanzara, N. L. Saini, M. Brunelli, F. Natali, A. Bianconi, 
 P. G. Radaelli, and S.-W. Cheong,
   Phys. Rev. Lett.  {\bf 81},                      
   878 (1998).                           

  \bibitem{lang-firsov}
 I. G. Lang and Yu. A. Firsov,
 Zh. Eksp. Teor. Fiz. {\bf 43}, 1843 (1962).

   \bibitem{sudhakar1}   
 Can be justified on the grounds that $\tau$ is independent of position.

   \bibitem{gosar}   
   P. Gosar,
   J. Phys. C: Solid State Phys. {\bf 8},                      
   3584 (1975).                          

   \bibitem{sudhakar2}   
We get the polaronic band energy  by replacing
$ g_{0}^2 $ with $\theta$ in  
 Eq.\ (\ref{Ek}).

   \bibitem{sudhakar3}   
 S. Yarlagadda, cond-mat {\bf 9809130}.

   \bibitem{Meq1}   
It should be noted that within the magnetic polarons 
$M \approx M_S$.

   \bibitem{khomskii}   
D. Khomskii, cond-mat {\bf 0004034}.

   \bibitem{sudhakar4}   
In Ref.\ \onlinecite{lee}, the authors
did not realize this fact.


\end{references}
  
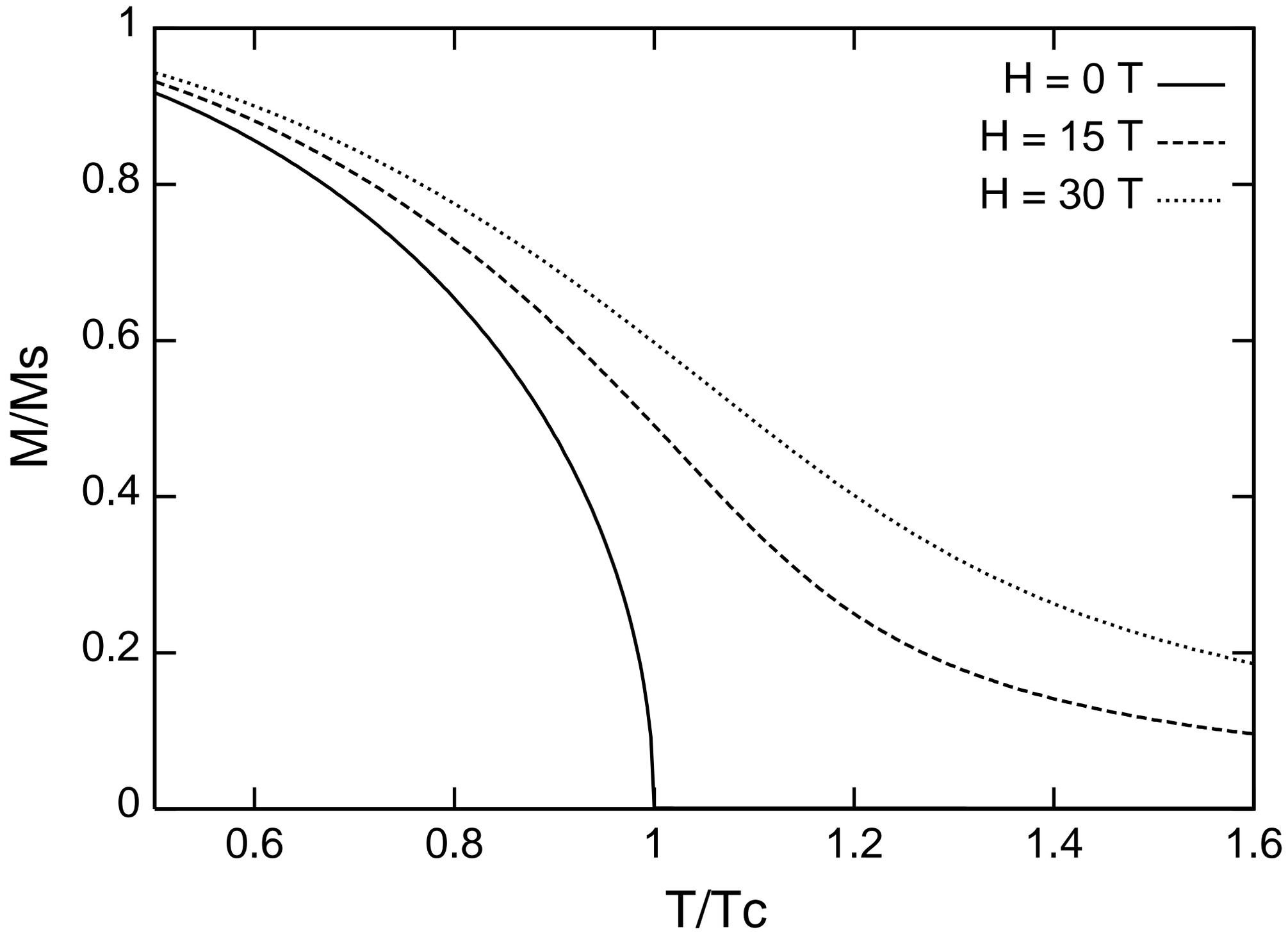
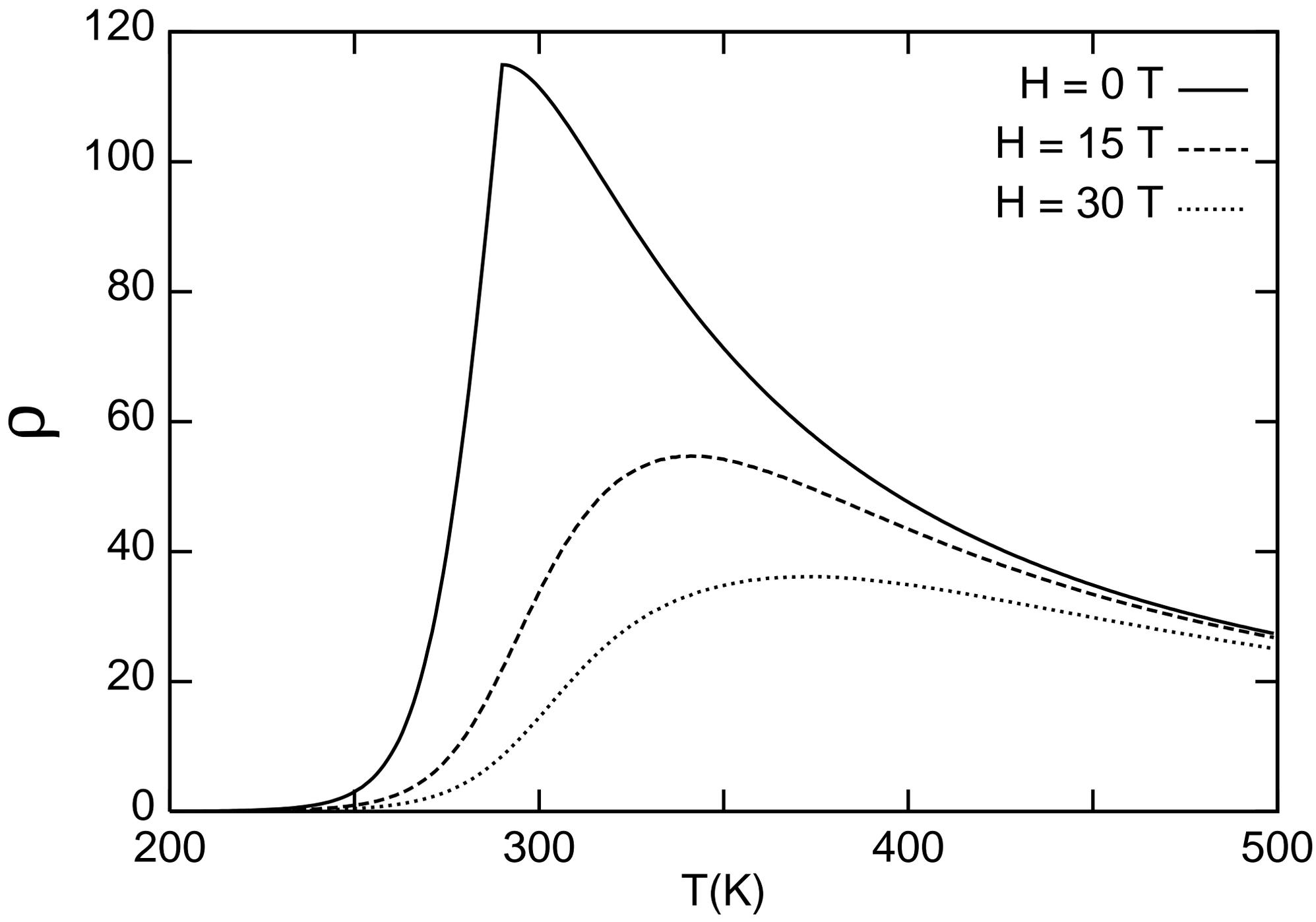
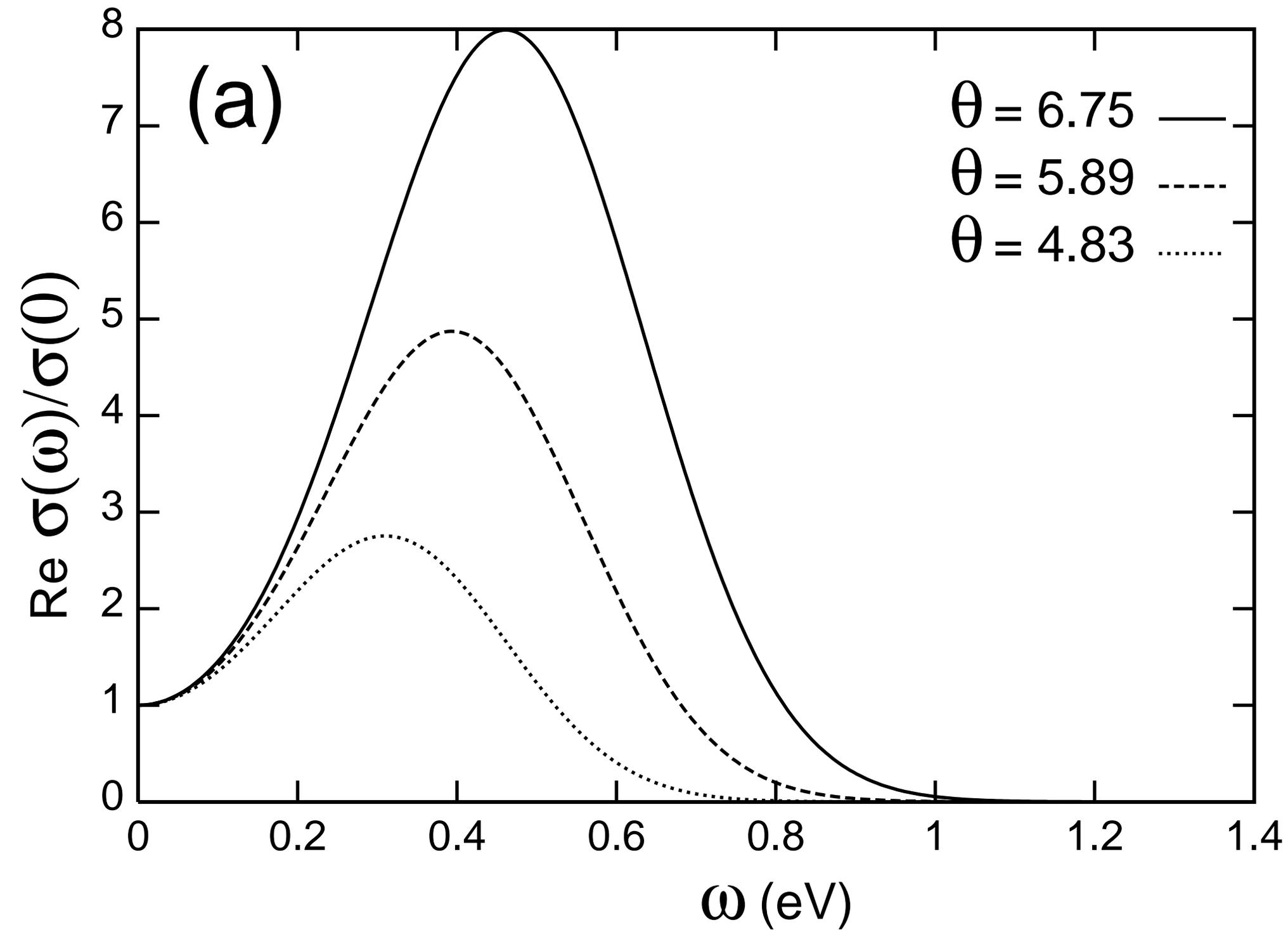
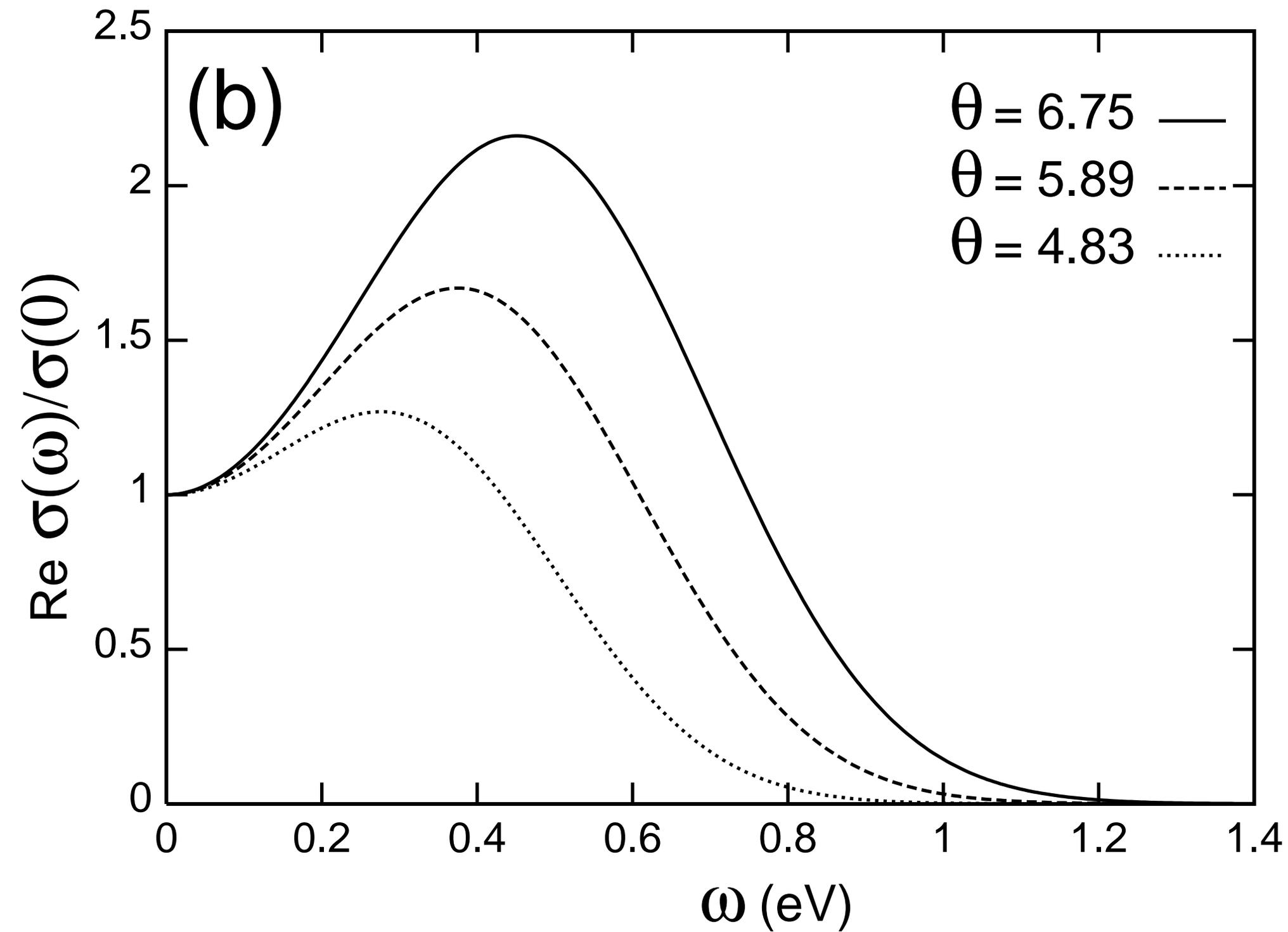
\end{document}